\documentclass[10pt]{iopart}
\usepackage{xcolor}
\definecolor{darkblue}{rgb}{0,0,.6}
\usepackage{braket}
\usepackage{amssymb}
\usepackage{cite}
\usepackage[colorlinks=true, citecolor=darkblue, linkcolor=darkblue, menucolor=darkblue, urlcolor=darkblue]{hyperref}
\usepackage{float}

\usepackage{graphicx}

\newcommand{\YbB}{$\mathrm{YbB_4}$}
\newcommand{\RB}{$\mathrm{REB_4}$}
\newcommand{\YbAlB}{$\mathrm{\alpha}$-$\mathrm{YbAlB_4}$}
\newcommand{\Ybtwo}{$\mathrm{Yb^{2+}}$}
\newcommand{\Ybthree}{$\mathrm{Yb^{3+}}$}
\newcommand{\dnh}{$\Delta n_h(0)$}
\newcommand{\dnht}{$\Delta n_h(T)$}
\begin{document}

\title{Intermediate Valence State in \YbB{} Revealed by Resonant X-ray Emission Spectroscopy}

\author{Felix Frontini$^{1,\dagger}$, Blair W. Lebert$^1$, K. K. Cho$^2$, M. S. Song$^2$, B. K. Cho$^2$, Christopher J. Pollock$^3$ and Young-June Kim$^{1,\ddagger}$}
\address{$^1$ Department of Physics, University of Toronto, 60 St. George Street, Toronto, ON, M5S 1A7, Canada}
\address{$^2$ School of Materials Science and Engineering, Gwangju institute of Science and Technology, 61005 Gwangju, Korea}
\address{$^3$ Cornell High Energy Synchrotron Source (CHESS), Cornell University, Ithaca, New York 14853, United States}

\ead{$^\dagger$ felix.frontini@mail.utoronto.ca,$^\ddagger$ youngjune.kim@utoronto.ca}

\date{\today}

\begin{abstract}
We report the temperature dependence of the Yb valence in the geometrically frustrated compound \YbB{} from 12 to 300 K using resonant X-ray emission spectroscopy at the Yb $L_{\alpha_1}$ transition. We find that the Yb valence, $v$, is hybridized between the $v=2$ and $v=3$ valence states, increasing from $v=2.61\pm0.01$ at 12 K to $v=2.67\pm0.01$ at 300 K, confirming that \YbB{} is a Kondo system in the intermediate valence regime.
This result indicates that the Kondo interaction in \YbB{} is substantial, and is likely to be the reason why \YbB{} does not order magnetically at low temperature, rather than this being an effect of geometric frustration. Furthermore, the zero-point valence of the system is extracted from our data and compared with other Kondo lattice systems. The zero-point valence seems to be weakly dependent on the Kondo temperature scale, but not on the valence change temperature scale $T_v$.
\end{abstract}
\ioptwocol
\section{\label{sec:intro}Introduction\protect}
\begin{figure}[!b]
\includegraphics[width=\columnwidth]{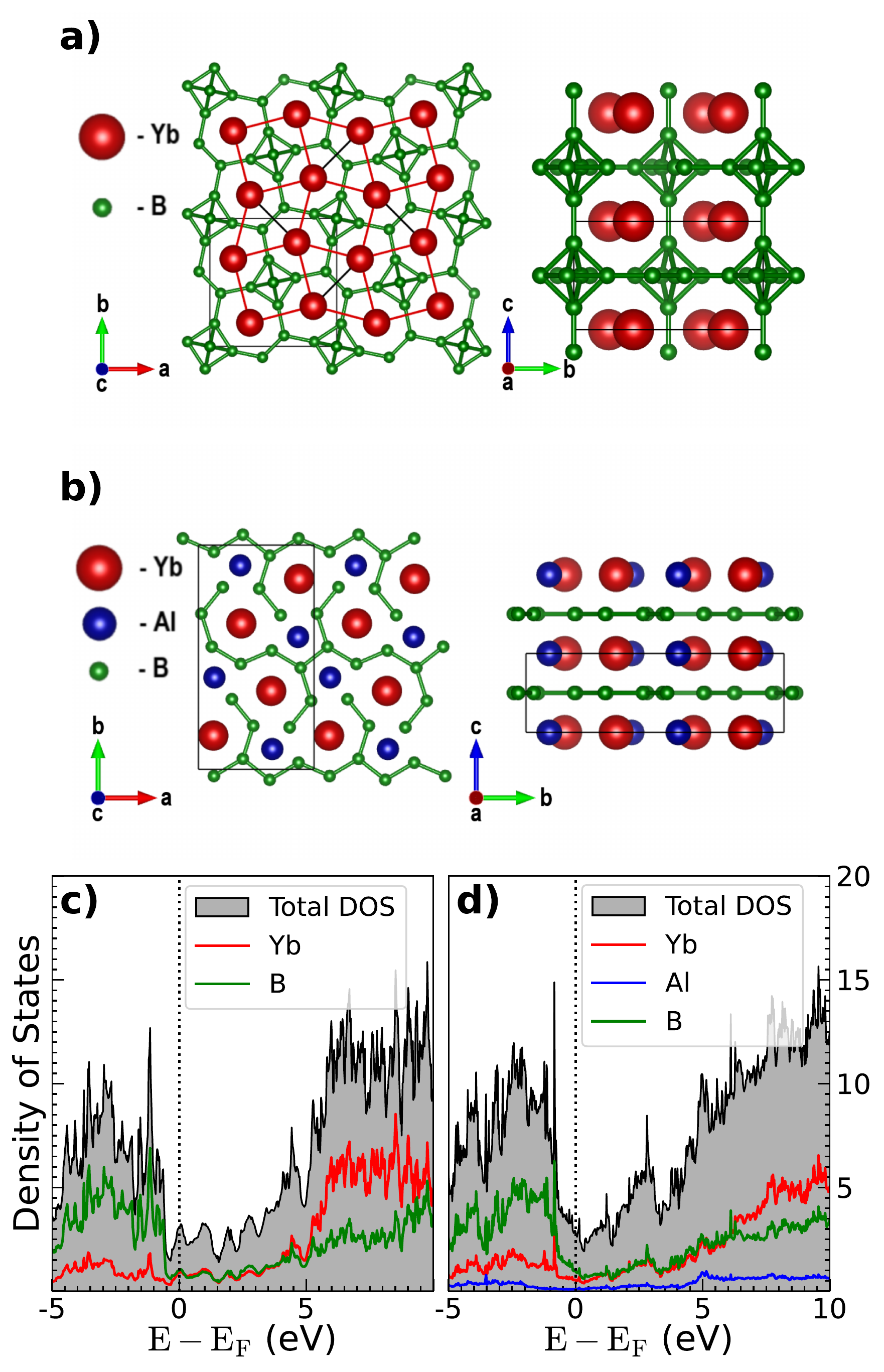}
\caption{a) \YbB{} crystal structure, viewed along the $c$ (left) and $a$ (right) axes. To illustrate the Shastry-Sutherland Lattice mapping, Yb-Yb nearest and next-nearest neighbour bonds are illustrated in red and black respectively. b) \YbAlB{} crystal structure, viewed along the $c$ (left) and $a$ (right) axes. c) \YbB{} \cite{MP_Jain2013,MPAPI,MPpymatgen,MPYbB4_Etourneau1979,MPYbB4_sd_0525202,MPYbB4_sd_0525235,MPYbB4_sd_0526220,MPYbB4_sd_1908283} and d) \YbAlB{} density of states from the Materials Project \cite{MP_Jain2013,MPAPI,MPpymatgen,MPYbAlB4_Derkhachenko1991,MPYbAlB4_Macaluso2007,MPYbAlB4_Mikhalenko1980}.}
\label{fig:fig1}
\end{figure}

In geometrically frustrated systems, frustration arises from the inability to minimize the total energy of competing antiferromagnetic interactions, resulting in a plethora of possible novel magnetic ground states \cite{geometric_frustration_general}. Perhaps the most well known example of this is the theoretical spin 1/2 Kagom\'{e} lattice whose ground state is truly disordered and continues to fluctuate in the absence of thermal energy, a so-called quantum spin liquid \cite{Kagome}. Another interesting real-world example of magnetic frustration is found in the family of Rare-Earth (RE) tetraborides $\mathrm{REB_4\ (RE=La,Ce,...,Yb,Lu)}$. The \RB{} family crystallizes in a tetragonal structure (space group $P4/mbm$, shown in Figure \ref{fig:fig1}.a) that can be mapped to the frustrated Shastry-Sutherland lattice within the \textit{ab} plane \cite{REB4_SSL_structure}. Frustration in the Shastry-Sutherland lattice is quantified by the relative strengths of the nearest and next-nearest neighbour exchange interactions $J$ and $J'$ \cite{SSL_frust}. For $J'/J\gtrsim 1$, the formation of simple N\'{e}el-type antiferromagnetic order is suppressed and novel magnetic properties are observed, such as the fractional magnetization plateaus observed in $\mathrm{NdB_4}$ and $\mathrm{HoB_4}$ \cite{NdHoB4}. When frustration is tuned higher ($J'/J\gtrsim 2$), long range magnetic order is entirely suppressed by the formation of a valence bond solid state. \YbB{} is unique among the $\mathrm{REB_4}$ compounds in that it does not show magnetic order down to 0.34 K, which may be explained by the magnetic frustration effects of the Shastry-Sutherland lattice \cite{MPYbB4_Etourneau1979}.\par 
Magnetic frustration is, however, not the only possible explanation for the lack of magnetic order in \YbB{}. Possibly responsible for this behaviour is the Kondo physics typical of many Yb-based systems, which has garnered significant interest over the years due to phenomena such as unconventional superconductivity and non-Fermi-liquid (NFL) behaviour \cite{Kondo_NFL_USC}. In such systems, physical properties are determined by the strength of an effective antiferromagnetic exchange between the $4f$ and conduction electrons dubbed the Kondo interaction, characterized by a system specific hybridization strength \cite{kondo_int,doniach}. When the Kondo interaction is much stronger than the intersite antiferromagnetic exchange, the $4f$ moments are screened from each other. This screening suppresses the formation of long range order and results in a system with strongly renormalized conduction electrons, a so-called Kondo lattice. When the hybridization strength is only moderate the result is a heavy fermion system but when it is large the result is an intermediate valence (IV) system, characterized by charge fluctuations and a strong physical hybridization of valence states \cite{IV1,IV2}. That is to say, an IV Yb-based system possesses an electron configuration that is quantum mechanically mixed between $[Xe]4f^{13}6s^25d^1$ (\Ybthree{}) and $\ket{[Xe]4f^{14} 6s^2}$ (\Ybtwo{}) configurations: $\ket{\psi} = \alpha \ket{ [Xe]4f^{13} 6s^2 5d^1} + \beta \ket{[Xe]4f^{14} 6s^2}$. Practically, the degree of hybridization is quantified by the average number of holes $n_h$ in the $4f$ band, which deviates from $n_h=1$ as hybridization strength increases.  This is related to overall valence simply by $v=2+n_h$, where $v$ is the Yb valence.\par
\YbB{} was first hypothesized to possess `abnormal valency' in 1972 based on abnormality of its lattice parameters, but has only recently been formally proposed as an IV metal based on profiles of the resistivity, in-plane magnetic susceptibility, and pressure dependence of the magnetic susceptibility \cite{YbB4_abnormal_lattice_constants,YbB4main,YbB4_IV}.
However, still to date no direct confirmation of intermediate valence through spectroscopic method has been carried out. In this paper, we seek to address this knowledge gap by directly characterizing the valence state in \YbB{} using X-ray spectroscopy. X-ray absorption spectroscopy (XAS) has proven to be an invaluable tool in studying valence hybridization in RE systems due to its relative experimental simplicity, and its nature as a truly bulk sensitive probe. In particular, this differentiates XAS from other techniques such as photo-emission spectroscopy (PES) which may only probe surface or near surface properties  \cite{PES1,PES2}. The primary limitation of XAS, however, is the precision with which absorption features can be resolved. With experiments performed at the RE $L_3$ edge ($2p^{3/2}\to 5d $), a standard in valence spectroscopy, the ability to resolve white line features is hampered by the large spectral width, a by-product of the short lifetime of the final state $2p$ core hole. The development of resonant X-ray emission spectroscopy (RXES) has significantly improved the resolution of such experiments, due to the significantly longer lifetime of the fluorescence final state core hole. In particular, high energy resolution fluorescence detected X-ray absorption spectroscopy (HERFD-XAS) measures X-ray absorption by monitoring the intensity of the fluorescence line. For an $L_3$ edge experiment, the $L_{\alpha_1}$ decay channel ($3d^{5/2}\to2p^{3/2}$) is typically monitored, putting the spectral width dependency on the lifetime of a $3d$ core-hole, which decreases the spectral width from $\sim$ 4.3 eV to $\sim$ 1.4 eV \cite{HERFD,atomic_widths}. \par

We have used RXES techniques to directly confirm that \YbB{} exists in an IV state at all measured temperatures, with a non-integer valence that increases from  $v=2.61\pm 0.01$ at 12 K to $v=2.67\pm0.01$ at 300 K. We compare the temperature scaling of the $4f$ hole occupancy in \YbB{} to other Yb-based Kondo systems and find that the universal scaling observed in weakly mixed-valent Kondo lattices does not extend to systems, \YbB{} included, in the IV regime \cite{kummer}. We instead find that there appears to be no constraints based on zero point valence of the temperature scale of valence change in IV systems. We also investigate the extension of the power law relationship observed between the zero point valence and Kondo temperature $T_K$ in weakly mixed-valent Kondo lattices to systems in the IV regime. A power law is found to be appropriate, however, in contrast to the previous scaling exponent of $n=2/3$, an exponent of $n\sim1/3$ is found to be more appropriate for a wider range of materials including IV compounds \cite{kummer}.

\begin{figure*}[!ht]

\includegraphics[width=\textwidth]{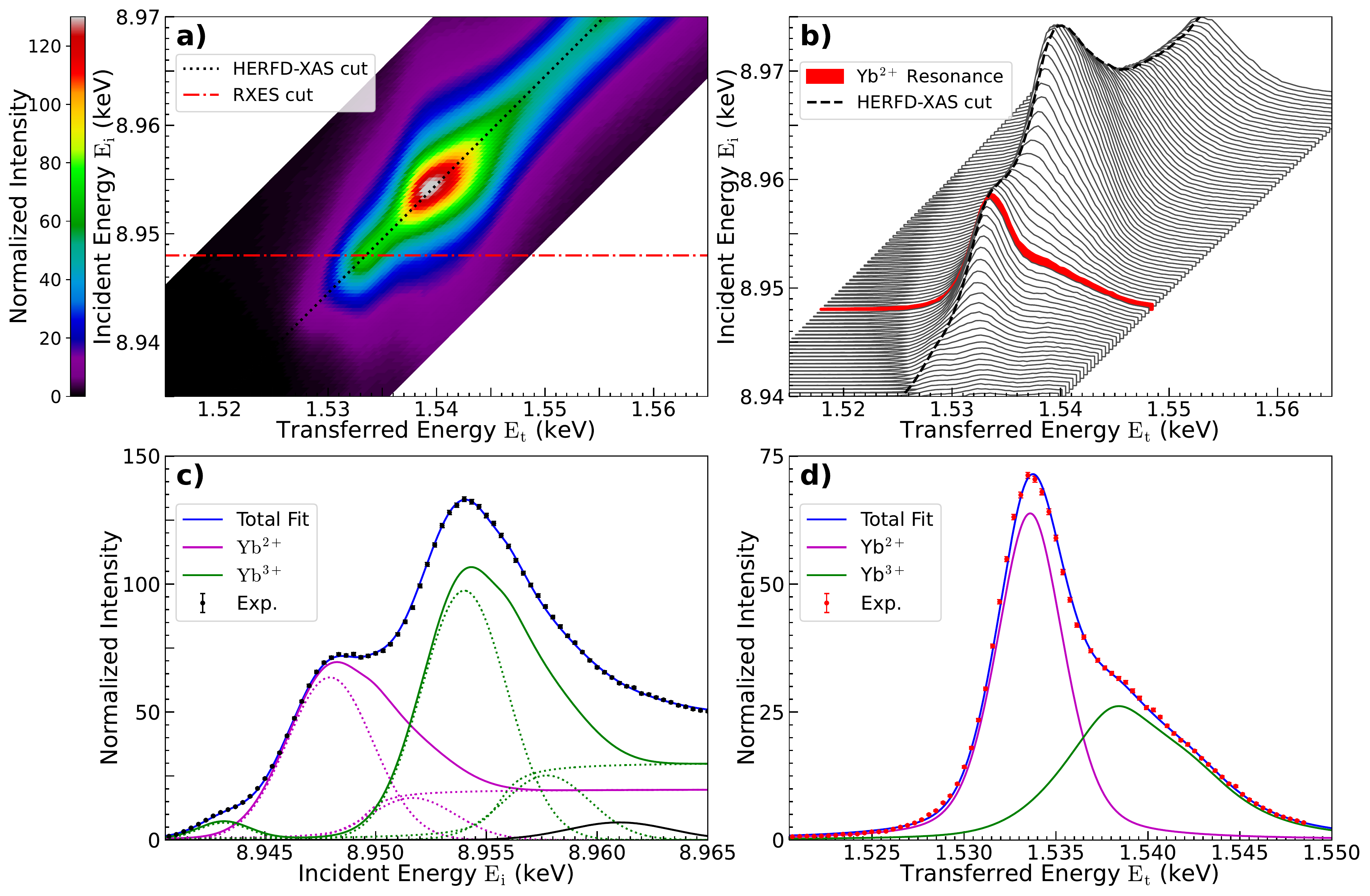}

\caption{\label{fig:fig2}a) \YbB{} $L_{\alpha_1}$ RXES colormap measured at 12 K with HERFD-XAS cut at $L_{\alpha_1}$ nominal and RXES cut at \Ybtwo{} resonance illustrated. b) RXES cuts as $E_i$ is tuned from 8.935 keV to 8.97 keV, with HERFD-XAS cut illustrated and \Ybtwo{} resonance highlighted. c) HERFD-XAS cut and corresponding fit at 12 K. d) RXES cut at \Ybtwo{} resonance and corresponding fit at 12 K.}
\end{figure*}

\section{\label{sec:experiment}Experimental Details\protect}
X-ray spectroscopy measurements were performed at the Cornell High Energy Synchrotron Source (CHESS) at the PIPOXS beamline. To prepare the sample, a \YbB{} single crystal was ground and diluted with BN powder to two absorption lengths just above the Yb $L_3$ edge. The experiment was performed at the Yb $L_3$ edge ($E_i$ = 8.9436 keV) while monitoring fluorescence in the $L_{\alpha_1}$ decay channel ($E_f\sim$ 7.415  keV). On the incident optics side, a cryogenically cooled Si(311) double crystal monochromator was used. Rhodium coated vertical and horizontal focusing mirrors set to 4 mrad were employed for the purposes of harmonic rejection. The beam was focused to a spot size of $\mathrm{100x400\ \mu m^2}$ at the sample, with a flux of $\mathrm{\sim 3.1 \times 10^{11}\ ph/s}$ at 8.980  keV. For the receiving optics, a Johann-type spectrometer equipped with spherically-bent Si(620) crystal analyzers (radius 0.85 m) and a Pilatus 100K solid-state detector were used. We scanned the incident X-ray energy range 8.935--8.970 keV at constant emitted X-ray energy and iterated while incrementing the selected emitted X-ray energy, scanning the emitted X-ray energy range of 7.3995--7.4301 keV to create a 2D RXES map of emission intensity. Transmission through a Yb reference foil was monitored in order to calibrate the incident energy. This was done by fitting the Yb reference absorption to determine the energy shift needed such that the absorption inflection point coincides with the Yb $L_3$ edge ($E_i=8.9436$ keV \cite{merritt}). Temperature dependence was studied from the base temperature of 12 K up to 300 K using a closed cycle refrigerator.
During the experiment, the beam spot was rastered every 5 incident energy scans ($\sim$ twenty minutes) to avoid beam damage. No sample degradation was observed in this time frame during initial testing. Due to manual packing of the powder, it was necessary to correct the data taken at different sample positions for concentration variations. Normalization factors to account for variations in concentration were obtained by tuning the incident X-ray energy to well above the $L_3$ absorption edge and measuring the intensity of fluorescence at each of the raster positions.
\section{\label{sec:results}Results}
\subsection{\label{sec:HERFD}HERFD-XAS}
In Figure \ref{fig:fig2}.a) we show the RXES scans at 12 K in a 2D pseudocolour map. The map is plotted against axes of incident X-ray energy $E_i$ along the vertical and transferred energy $E_t$ along the horizontal, where $E_t=E_i-E_f$ is simply the difference between incident and emitted X-ray energies. In this plot the diagonal corresponds to cuts with constant emitted X-ray energy $E_f$. 
The highlighted diagonal corresponds to the monitored X-ray fluorescence at the nominal $L_{\alpha_1}$ transition (X-rays of constant emitted energy $E_f\sim$ 7.4145 keV) measured as incident energy is scanned. This is the nominal HERFD-XAS spectrum, which is shown in greater detail in Figure \ref{fig:fig2}.c). The HERFD-XAS spectrum is particularly useful since it provides the true relative population of the \Ybtwo{} and \Ybthree{} states, given by the ratio of the integrated intensities of the respective components, while providing a large advantage in energy resolution compared to standard XAS. The HERFD-XAS spectrum was curve fitted to extract the intensity of the \Ybtwo{} and \Ybthree{} components. 
Figure \ref{fig:fig2}.c) shows the individual contributions of the \Ybtwo{} and \Ybthree{} lineshapes, the latter of which lies higher in energy due to the additional repulsive coulomb interaction between the intermediate state $2p^{3/2}$ core hole and the \Ybthree{} $4f$ hole that is not present in the \Ybtwo{} state. The \Ybtwo{} lineshape consists of a peak to fit the $L_3$ edge white line and an arctangent step function to fit the background above-edge absorption to continuum. We find that quality of fit necessitates the use of a doublet peak, which we fit with two Gaussian functions separated in energy by $\sim$ 3.7 eV, with the first peak centered at 8.948 keV. This splitting is a common feature in the absorption spectra of Yb based IV materials and is attributed to crystal field splitting of the $5d$ levels, an explanation we believe to be applicable in our case as there is evidence of large crystal field effects in \YbB{} \cite{CEF_exp,CEF_model,YbB4main}. The \Ybthree{} component is fitted with the same lineshape shifted in energy by $\sim$ 6 eV, with the addition of a peak in the pre-edge region. This pre-edge peak has previously been attributed to the dipole forbidden quadrupolar $2p_{3/2}\to 4f$ transition from the core hole to the $\mathrm{Yb^{3+}} 4f$ vacancy \cite{YbCu2Si2_valence}. The intensity of the \Ybtwo{} feature is given by the integrated intensity of the doublet peak, while the intensity of the \Ybthree{} feature is given by the integrated intensity of the doublet and pre-edge peaks. In both cases the intensity due to the background step function is not considered. We also observe a feature at high energy which is fitted with a Gaussian function peaked at 8.962 keV. This is attributed to normal fluorescence and is fitted solely for aesthetic purposes; its removal does not affect our results.\par
\begin{figure}[!b]
    \centering
    \includegraphics[width=\columnwidth]{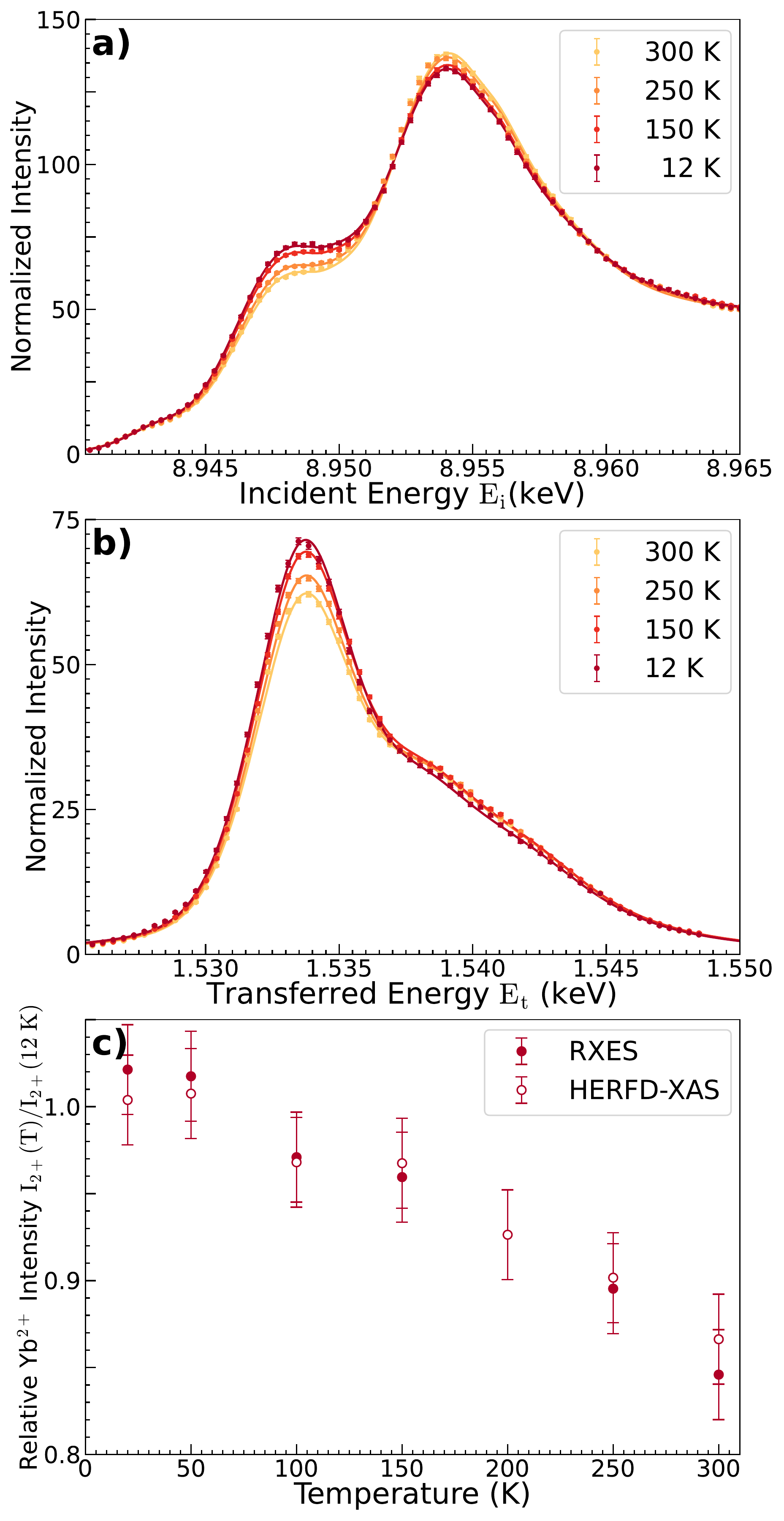}

    \caption{ Temperature dependence between 12 K and 300 K of a) the HERFD-XAS spectra, b) the RXES spectra at \Ybtwo{} resonance, and c) the relative \Ybtwo{} intensity, $\mathrm{I_{2+}(T)/I_{2+}(12\ K)}$, measured by both HERFD-XAS and RXES.}
    \label{fig:fig3}
\end{figure}

Immediately evident from our fit of the HERFD-XAS spectra at 12 K is the existence of an intermediate, or non-integer, valence state in \YbB{}, so evident by the fact that we fit non-zero intensities for the lineshapes corresponding to both \Ybtwo{} and \Ybthree{} valence states. This suggests that the lack of magnetic order in \YbB{} is attributable to the dominance of the Kondo effect. We extend this fitting procedure to the HERFD-XAS spectra at all measured temperatures to quantify any temperature dependence. In Figure \ref{fig:fig3}.a) we show fitted HERFD-XAS spectra measured between 12 K and 300 K. We visually observe that the \Ybtwo{} feature is suppressed as temperature increases while the \Ybthree{} is enhanced. This is less obvious with the pre-edge \Ybthree{} peak due to the reduced scale of the feature and bleed from the tail of the \Ybtwo{} peak which scales oppositely. To quantify these changes we examine the temperature dependence of the \Ybtwo{} feature intensity compared to its measured intensity at 12 K, illustrated in Figure \ref{fig:fig3}.c). We find that the \Ybtwo{} feature intensity decreases to $\sim 85$\% of the 12 K value by 300 K. This is a clear indication that the \Ybtwo{} occupancy decreases with temperature as expected for a Kondo system \cite{AIM_valence,YbAgCu4valence}.
\subsection{\label{sec:RXES}RXES}
By monitoring X-ray fluorescence while scanning emitted X-ray energy and with incident energy tuned to \Ybtwo{} resonance we obtain the desired RXES spectrum, illustrated in Figure \ref{fig:fig2}.d) for the measurement at 12 K. \Ybtwo{} resonance is here defined as the incident energy point closest to the \Ybtwo{} peak energy from our HERFD-XAS fit ($E_i=$8.948 keV). From our total 2D RXES map, this is taken as a single incident energy cut, as illustrated in Figure \ref{fig:fig2}.a). Though the RXES spectrum does not provide the direct equivalency between intensity and fractional population that HERFD-XAS does, it provides us with the ability to selectively enhance the intensity of the weaker \Ybtwo{} component, as illustrated in Figure \ref{fig:fig2}.b). This enhancement in intensity aids in more accurately tracking the temperature scaling of the \Ybtwo{} component intensity and is often critical for systems with smaller \Ybtwo{} contributions \cite{kummer}. In the case of \YbB{}, however, the \Ybtwo{} component is strong enough to be well resolved in the HERFD-XAS spectrum. Instead, RXES is used to validate the HERFD-XAS results and provide a quantitative metric for estimating the uncertainty in our valence estimates.\par
\begin{figure}[!b]
    \centering
    \includegraphics[width=\columnwidth]{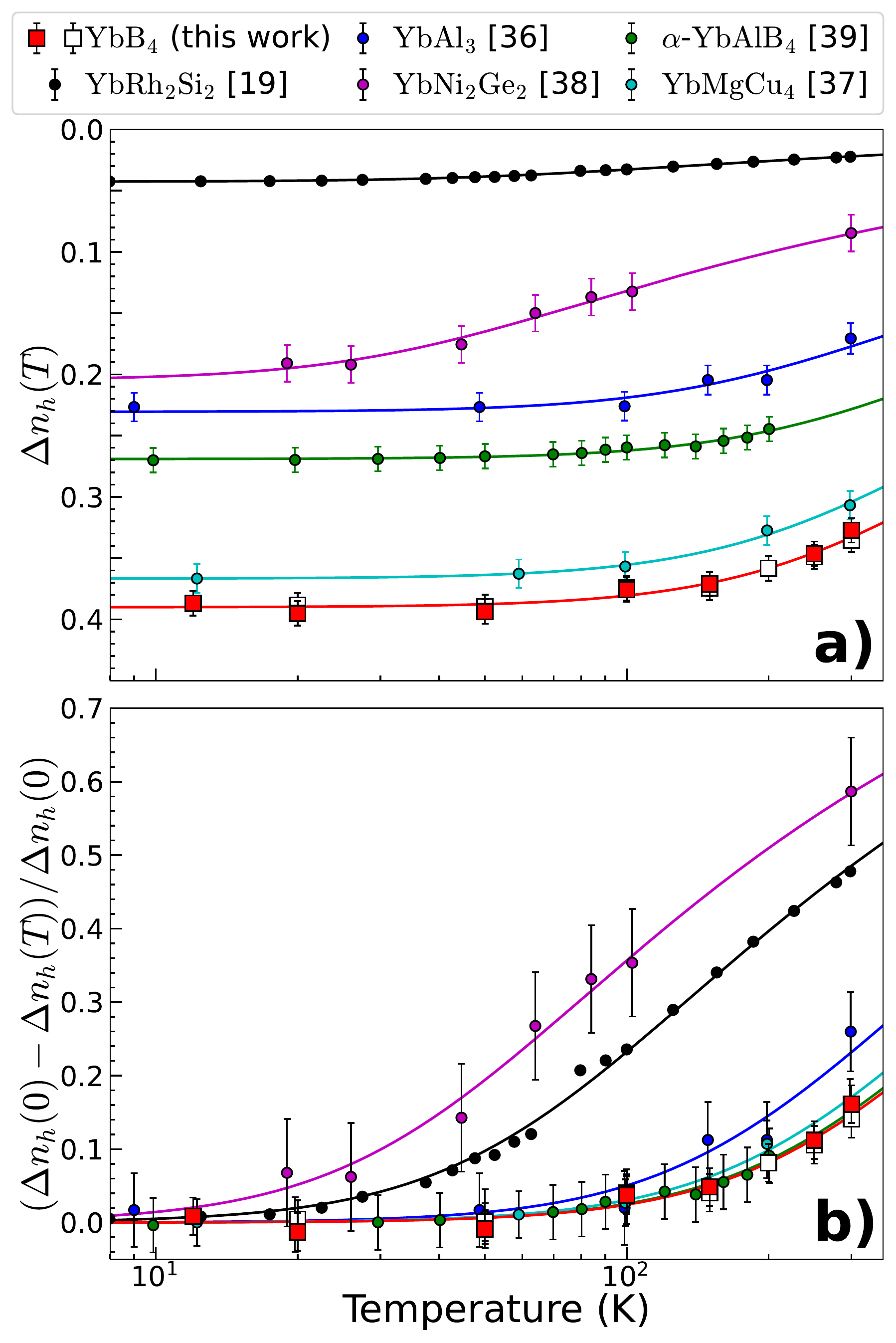}

    \caption{Temperature dependence of the \YbB{} $4f$ hole occupancy in a) absolute values and b) normalized to the low temperature value. Open (closed) squares indicate values obtained from RXES (HERFD-XAS). Scaling of other Yb-based Kondo lattices shown for comparison, with data reproduced from refs  \cite{YbAl3valence,YbXCu4valence,kummer,YbNi2Ge2_valence,YbAlB4_valence}.}
    \label{fig:fig4}
\end{figure}
In order to accurately determine the change in the \Ybtwo{} component intensity with temperature it is necessary to possess a consistent total emission intensity scale at all temperatures. We found that the scaling performed to correct for variations in concentration was insufficient for this purpose and that a second normalization was required. To this end, we normalize by the summed integrated intensities of the \Ybtwo{} and \Ybthree{} HERFD-XAS features, such that the total integrated intensity of the HERFD-XAS is scaled to unity at every temperature. This also simplifies further analysis in that the integrated intensity of each HERFD-XAS feature is now directly equal to the fractional occupation of the corresponding electronic state.
Following this normalization the RXES spectrum was fitted to extract the individual \Ybtwo{} and \Ybthree{} lineshapes. Figure \ref{fig:fig2}.d) shows the individual contributions from the \Ybtwo{} and \Ybthree{} lineshapes, the latter of which lies higher in energy due to the additional repulsive coulomb interaction between the final state $3d^{5/2}$ core hole and the \Ybthree{} $4f$ hole. The \Ybtwo{} lineshape is comprised of a single pseudo-Voigt function peaked at 1.532 keV. The \Ybthree{} lineshape is composed of a pseudo-Voigt function peaked at 1.538 keV with an asymmetric pseudo-Voigt function to fit the extended fluorescence tail. As with our fit of the HERFD-XAS spectrum, our fit of the RXES spectrum indicates an intermediate valence state in \YbB{}. We extend this fitting procedure to all temperatures and investigate the temperature dependence of the RXES spectrum, as shown in Figure \ref{fig:fig3}.b). We observe that, as in the HERFD-XAS spectrum, the \Ybtwo{} feature is suppressed as temperature increases from 12 K to 300 K. A slight enhancement of the \Ybthree{} feature is also observed with increasing temperature, however this change is not large due to the fact that we are off-resonance of the \Ybthree{} feature. These changes are quantified in the same way as the HERFD-XAS spectrum, by examining the temperature evolution of the \Ybtwo{} feature, shown alongside the HERFD-XAS estimate in Figure \ref{fig:fig3}.c). Using RXES we observe the same reduction of $\sim 15\%$ in the \Ybtwo{} feature intensity between 12 K and 300 K.
\section{\label{discussion}Discussion}
\begin{figure*}[!t]
    \centering
    \includegraphics[width=\textwidth]{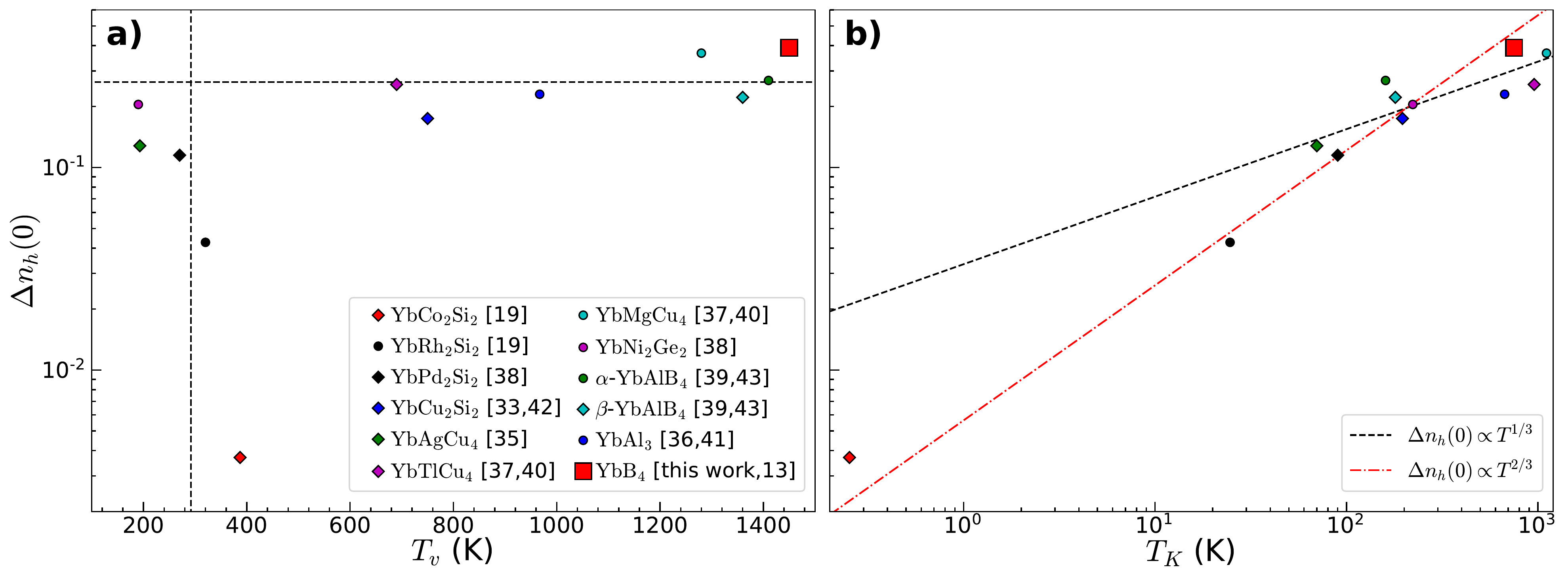}

    \caption{\label{fig:fig5}\dnh{} plotted against a) $T_v$ and b) $T_K$. Dashed lines in a) correspond to the average $T_v$ of materials with $\Delta n_h(0)<0.15$ and the average \dnh{} of materials with $\Delta n_h(0)>0.15$. The dashed line in b) corresponds to the power law $\Delta n_h(0) \propto T_K^{1/3}$, while the dash dotted line corresponds to the power law $\Delta n_h(0) \propto T_K^{2/3}$. $T_v$ obtained by fitting the data from refs  \cite{YbXCu4valence,YbAgCu4valence,YbAl3valence,YbAlB4_valence,YbCu2Si2_valence,YbNi2Ge2_valence,kummer} to \Eref{eq:1}. $T_K$ reproduced from refs  \cite{YbB4main,YbXCu4_Tk,kummer,YbAgCu4valence,YbNi2Ge2_valence,YbAl3_Tk,YbCu2Si2_Tk,YbAlB4_Tk}.}
\end{figure*}
Our experimental results confirm that \YbB{} is in an IV state at all temperatures, a clear indication that the Kondo interaction in \YbB{} is strong and is therefore likely responsible for the displayed lack of magnetic order. Our results also show that the intensity of the \Ybtwo{} component in both HERFD-XAS and RXES spectra is temperature dependent, with it decreasing in intensity by $\sim$ 15\% between 12 K and 300 K. While the relative change in \Ybtwo{} intensity is interesting, we are more interested in the absolute change and overall scale of the Yb valence in \YbB{}. We calculate this by proxy of the deviation in the $4f$ hole occupancy from $n_h=1$, which we notate as $\Delta n_h = 1-n_h$, and which is related to the valence by $v=3-\Delta n_h$. Physically, $\Delta n_h$ represents the fractional occupation of the \Ybtwo{} state. We determined $\Delta n_h$ using the intensity of the \Ybtwo{} and \Ybthree{} features at each temperature, shown in Figure \ref{fig:fig4}. alongside several other Yb based Kondo lattice systems for comparison.
In the HERFD-XAS spectra, \dnht{} is simply equal to the normalized integrated intensity of the \Ybtwo{} peak at temperature T, $I_{2+}(T)$. 
For the RXES spectra, because of its resonant nature, we adopt a starting value from HERFD-XAS. We adopt the value at 12 K where the \Ybtwo{} component is largest, and hence where the HERFD-XAS fit is the most reliable.
From this point, the RXES estimation of \dnht{} is given by this value scaled by the change in the RXES spectrum \Ybtwo{} intensity relative to its value at 12 K
\begin{equation*}
\centering
        \Delta n_h(T) =I_{2+}(T)\cdot\frac{\Delta n_h(12\ \mathrm{K}) }{I_{2+}(12\ \mathrm{K})}
\end{equation*}
Our calculations reveal that \dnht{} decreases monotonically from $0.39\pm 0.01$ at 12 K to $0.33\pm0.01$ at 300 K. 
This translates to a change in valence from $v=2.61\pm0.01$ to $v=2.67\pm0.01$.\par
To understand the temperature dependence of Yb valence, we employ the method of Kummer et al. to estimate the zero point \Ybtwo{} occupancy, \dnh{}, and the characteristic valence change temperature $T_v$ \cite{kummer,valence_fit_param}.
Here $T_v$ simply corresponds to the temperature at which $\Delta n_h(T)=1/2 \Delta n_h(0)$. These are estimated by an empirical model which takes the form
\begin{equation}
    \label{eq:1}
    \Delta n_h(T)=\Delta n_h(0)/[1+(2^{1/0.21}-1)(T/T_v)^2]^{0.21}
\end{equation}
The empirical form of \Eref{eq:1} fits our data well, and returns a zero point \Ybtwo{} occupancy of \dnh{}= $0.390\pm0.003$, as well as a valence change temperature $T_v$ on the order of 1500 K.
Comparison to other Yb-based Kondo systems shows qualitatively similar temperature scaling of the \Ybtwo{} occupancy amongst all the compounds, however, the normalized scaling in Figure \ref{fig:fig4}.b) illustrates some notable differences.
In particular, it appears that the scaling of the \Ybtwo{} occupancy is slower for systems that are more strongly mixed-valent (i.e. have larger \dnh{}).\par
To quantify this, we have compared the fitted values of $T_v$ and \dnh{} of various Yb-based Kondo systems, illustrated in Figure \ref{fig:fig5}.a). 
In Figure \ref{fig:fig5}.b) we plot \dnh{} vs. the reported Kondo temperatures $T_K$ available from literature of the selected compounds.
Note that all values of $T_v$ and \dnh{} are obtained using \Eref{eq:1} to provide consistency, and that while other choices of model give different values for $T_v$, the hierarchy of scaling between compounds is quite robust. 
As such, we put little stock in the specific values of $T_v$ but rather focus on the differences between compounds. 
As illustrated in Figure \ref{fig:fig5}, though there appears to be a rough scaling relationship between \dnh{} and $T_K$, there does not appear to be such a relationship between \dnh{} and $T_v$.
The scaling relationship between \dnh{} and $T_K$ is fitted to a power law of the form $\Delta n_h(0)\propto a T_K^n$, with $a\sim 1/30$ and $n\sim 1/3$. This exponent is quite different from the exponent of $n \sim 2/3$ found by Kummer et al., illustrated in Figure \ref{fig:fig5}b) for reference, a difference we attribute to choice of reference materials used for comparison, in particular the fact that we choose to focus on IV compounds (here phenomenologically defined as $\Delta n_h(0) \gtrsim 0.15$) whereas such materials are excluded from the analysis of Kummer et al. \cite{kummer}. Agreed upon by both our analysis and that of Kummer et al. is the existence of a sub-linear scaling relationship between \dnh{} and $T_K$. As noted by Kummer et al., it is expected that the precise nature of this relationship is complex, but is in agreement with a sub-linear relationship.\par
In Figure \ref{fig:fig5}.a), we illustrate that for materials that are only moderately mixed-valent ($\Delta n_h(0) \lesssim 0.15$), it appears that the valence change temperature $T_v$ is more or less fixed, on the order of $\sim 300$ K, as was shown by Kummer et al. \cite{kummer}. On the other hand, for materials in the IV regime this breaks down and instead it appears as if there are no constraints on $T_v$, with values ranging over more than an order of magnitude. This suggests that $T_v$ for IV systems is neither universal nor proportional to the strength of the Kondo interaction, and may instead be influenced in a more complex manner by external variables. This view is reinforced by a deeper comparison of \YbB{} and its closest analog in terms of both $T_v$ and \dnh{}, \YbAlB{}.\par
Both \YbAlB{} and \YbB{} possess an enhanced mass Fermi-liquid ground state and both systems have similar magnetic susceptibility profiles, with strong Ising anisotropy along the $c$-axis \cite{YbB4main,YbAlB4_FL,MPYbAlB4_Macaluso2007}. The crystal structure of both systems are shown in Figure \ref{fig:fig1}.b) for comparison. \YbAlB{} possesses an orthorhombic crystal structure (space group \textit{Pbam}), with layers of Yb-Al and B atoms stacked alternately along the crystal $c$-axis, and is visually similar to \YbB{}. In the Yb atom local environment of both materials there are striking similarities. In both materials, the B atoms are arranged in two overlapping heptagonal rings above and below the Yb atom along the $c$-axis. Furthermore, the in-plane Yb-Yb nearest and next-nearest neighbour separations are 3.740 and 3.777 \AA{} in \YbAlB{} compared to 3.681 and 3.720 \AA{} in \YbB{} \cite{MPYbB4_sd_0525202,MPYbAlB4_Mikhalenko1980}. Electronic density of states calculations shown in Figure \ref{fig:fig1}.c) and d) also show similarities between \YbAlB{} and \YbB{}. By DOS calculations both are metallic. Both show stronger contributions from B atom electrons below the Fermi level, close to equal Yb and B contributions at the Fermi level, and stronger Yb contributions above the Fermi level. Notably, in the case of \YbAlB{}, the Al atom electrons do not contribute strongly to the DOS in general and in particular at the Fermi level. Cumulatively, the similarity of these two materials suggest that some or all of crystal anisotropy, Yb atom local environment, and global electronic structure may have a significant effect on \dnh{} and $T_v$ in IV systems. Beyond the comparison of \YbB{} and \YbAlB{}, we point to the fact that the $\alpha$ and $\beta$ phases of $\mathrm{YbAlB_4}$ share each other as nearest materials in the \dnh{}-$T_v$ phase space to further reinforce this point.
\section{\label{conclusion}Conclusion}
We have studied the Yb valence in \YbB{} as a function of temperature between 12 K and 300 K using HERFD-XAS and RXES at the Yb $L_{\alpha_1}$ transition.
We find that the Yb valence is physically hybridized between the \Ybtwo{} and \Ybthree{} valence states, increasing monotonically from $v=2.61\pm0.01$ at 12 K to $v=2.67\pm0.01$ at 300 K. 
We are thus able to confirm that \YbB{} is a Kondo system with hybridization strength sufficient to place the system in the IV regime at all temperatures. This is indicative that the Kondo interaction in \YbB{} is significant and therefore is likely the mechanism preventing the formation of magnetic order in \YbB{}, suggesting that long range order is suppressed by the dominance of the Kondo interaction rather than by the effects of geometric frustration.
We also estimate the zero point \Ybtwo{} occupancy \dnh{} and temperature scale of valence change $T_v$ in \YbB{}, finding \dnh{} $\sim 0.39$ and $T_v\sim 1500$ K.
We compared this temperature scale with that of other Yb-based Kondo lattices. We observe that \dnh{} and the reported Kondo temperatures $T_K$ seem to be weakly correlated with a power law relationship with an exponent of $n\sim1/3$.
On the other hand, we find no correlation between $T_v$ and \dnh{} for systems in the IV regime, indicating that $T_v$ depends significantly on variables other than the Kondo interaction strength.
Anecdotal evidence based on similarities between \YbB{} and \YbAlB{} suggests that local environment, crystal anisotropy, and electronic structure may be among the external variables affecting $T_v$.\par
\ack
Work at the University of Toronto was supported by the Natural Sciences and Engineering Research Council (NSERC) of
Canada through the Discovery Grant No. RGPIN-2019-06449, Canada Foundation for Innovation, and Ontario Research Fund.
B.W.L. acknowledges support from NSERC (funding reference number PDF-546035-2020). 
This work was supported by the National Research Foundation of Korea (NRF) grant funded by the Ministry of Science and ICT (No. NRF-2022R1A2C1009516).
This work is based upon research conducted at the Center for High Energy X-ray Sciences (CHEXS) which is supported by the National Science Foundation under award DMR-1829070.
\section*{References}
\bibliographystyle{iopart-num}
\bibliography{apssamp}



\end{document}